\documentclass[preprint,groupedaddress,floatfix,nofootinbib]{revtex4}

\usepackage{euscript}
\usepackage{dcolumn}
\usepackage{graphicx}
\usepackage{epsfig}
\usepackage{hyperref}
\usepackage{graphicx}
\usepackage{dcolumn}
\usepackage{longtable}
\usepackage{times}
\usepackage{amsmath,amssymb,bm}
\usepackage[english]{babel}
\usepackage[latin1]{inputenc}
\usepackage{times}
\usepackage[T1]{fontenc}
\usepackage{color}
\usepackage{fancybox}
\usepackage{sidecap}
\usepackage{graphicx}
\usepackage{epsfig}
\usepackage{psfrag}
\usepackage{bbm}
\usepackage[english]{babel}
\usepackage[latin1]{inputenc}
\usepackage{times}
\usepackage[T1]{fontenc}
\usepackage{color}
\usepackage{fancybox}
\usepackage{sidecap}
\usepackage{psfrag}
\usepackage{bbm}
\usepackage{wasysym}
\usepackage{slashed}
\usepackage{tikz}
\usepackage{euscript}
\usepackage{booktabs}
\usepackage{algorithm}
\usepackage{algpseudocode}
\usepackage{float}

\begin{document}

\title{From Maximum Cut to Maximum Independent Set}
\author{Chuixiong Wu}
\author{Jianan Wang}
\author{Fen Zuo\footnote{Email: \textsf{zuofen@miqroera.com}}}
\affiliation{Hefei MiQro Era Digital Technology Co. Ltd., Hefei, China}

\begin{abstract}

The Maximum Cut (Max-Cut) problem could be naturally expressed either in a Quadratic Unconstrained Binary Optimization (QUBO) formulation, or as an Ising model. It has long been known that the Maximum Independent Set (MIS) problem could also be related to a specific Ising model. Therefore, it would be natural to attack MIS with various Max-Cut/Ising solvers. It turns out that this strategy greatly improves the approximation for the independence number of random Erd\H{o}s-R\'{e}nyi graphs. It also exhibits perfect performance on a benchmark arising from coding theory. These results pave the way for further development of approximate quantum algorithms on MIS, and specifically on the corresponding coding problems.

\end{abstract}
 \maketitle

\tableofcontents

\section{Introduction}

The relation between the Maximum Cut problem (Max-Cut) in graph theory and the physical Ising model could probably be traced back to the foundational paper \cite{Barahona-1982}. Knowledge on Max-Cut is of great help for our understanding of Ising model in various aspects \cite{Barahona-1982,Welsh-1993}. In turn, we could now use various physical Ising devices or simulations to solve difficult Max-Cut problems \cite{Annealing-2022}.

In the very same paper \cite{Barahona-1982}, the author also relate the Maximum Independent Set (MIS) problem to a specific Ising case. While Max-Cut with nonnegative edge weights could be well approximated with the famous Goemans-Williamson (GW) algorithms \cite{GW-1995}, MIS turns out to extremely difficult to approximate \cite{Hastad-1999}. A practical strategy would then be, first turn MIS into some form of Ising model, and then solve it with various Max-Cut/Ising solvers. This would be the approach pursued in this paper.

We arrange the remaining of the paper as follows. In the next section we introduce the Max-Cut problem, and elaborate its formulations as a Quadratic Unconstrained Binary Optimization (QUBO) problem and an Ising model. Some popular Max-Cut/Ising solvers will be discussed, and tested in a specific case: the Sherrington-Kirkpatrick (SK) model \cite{SK-1975}. Section III would be devoted to MIS. We start from its modeling with an Ising Hamiltonian, and then test the performance of various solvers. The focus would be on the random Erd\H{o}s-R\'{e}nyi (ER) graphs, and a specific benchmark arising from coding theory. In the last section we summarize and discuss possible generalizations.

\section{QUBO, Ising, and Maximum Cut}

Many realistic problems could be properly modeled by linear systems, and then solved in the framework of linear programming~\cite{Dantzig-1963}. A typical example is the maximum flow problem, and its linear-programming dual gives rise to the Minimum Cut problem~\cite{Cook-CO}. Nevertheless, it seems not easy to give the Max-Cut problem a natural linear-programming description. Instead, one could consider more general approaches, for example quadratic programming. Restricted to binary variables, one then considers the so-called QUBO framework. As the name suggests, the object function could be expressed in a quadratic form:
 \begin{equation}
 H(\mathbf{x})\equiv \mathbf{x}^{\rm T} \mathbf{Q} \mathbf{x},
 \end{equation}
where $\mathbf{x}=(x_i)\in \{0,1\}^{\otimes n}$ is an $n$-dimensional binary vector. Without loss of generality, we may assume that our goal is to minimize the function $H(\mathbf{x})$, and the matrix $\mathbf{Q}$ is symmetric.

Instead of binary variables, in physics people are more used to spin variables, which take values in $\{-1,+1\}$. The two could be related by
\begin{equation}
z=1-2x.
\end{equation}
With this relation, the QUBO cost function $ H(\mathbf{x})$ could be transformed to an Ising Hamiltonian:
\begin{equation}
H(\mathbf{z})=h_0-\sum_{i<j}J_{ij}z_iz_j-\sum_ih_iz_i,\label{eq.Ising}
\end{equation}
with
\begin{eqnarray}
h_0&=&\frac{1}{2}\sum_{i\le j}Q_{ij}\nonumber\\
h_i&=&\frac{1}{2}\sum_j Q_{ij}\nonumber\\
J_{ij}&=&-\frac{1}{2}Q_{ij},\quad (i\ne j).\label{eq.QUBO-Ising}
\end{eqnarray}
Therefore, a general QUBO problem is equivalent to an Ising model with local magnetic fields $h_i$.

Both the QUBO and Ising formulation could be used to describe Max-Cut in a natural way, and vice versa \cite{MQLib}.
For example, we could encode the above Ising Hamiltonian into a graph cut according to the following procedure. For each index $i$ we create a vertex $i$, and for each nonzero coupling $J_{ij}$ we create an edge $ij$. This gives rise to the underlying graph $G$ of the Ising model. To treat the coupling $J_{ij}$ and the magnetic fields $h_i$ in a uniform way, we could enlarge $G$ by including an auxiliary vertex $0$, and connect the new vertex with all the other vertices. We call the enlarged graph $G^*$. The couplings on the new edges are set as $J_{0i}=J_{i0}=h_i$. We use symbols with a bar, such as $\bar i$, to label the vertices in $G^*$. Now the Ising Hamiltonian could be formally written as
\begin{equation}
H(\mathbf{z})=h_0-\sum_{\bar i<\bar j}J_{\bar i \bar j}z_{\bar i} z_{\bar j}.
\end{equation}
Note that in doing so we have secretly made the choice $z_0=+1$.  An Ising state $|z_0,z_1,...,z_n\rangle$ naturally gives rise to a cut $V^*=V_1\biguplus V_2$ in $G^*$ as:
\begin{equation}
V_1=\{\bar i|z_{\bar i}=+1\}, \quad V_2=\{\bar i|z_{\bar i}=-1\}.
\end{equation}
If we further set the edge weights in $G^*$ as $w_{\bar i\bar j}=-2J_{\bar i\bar j}$ \cite{Barahona-1988},  then the weight of a general cut is given by
\begin{eqnarray}
w(\mathbf{z})&=& - \frac{1}{2}\sum_{\bar i<\bar j}w_{\bar i \bar j}z_{\bar i}z_{\bar j}+\frac{1}{2} w_0,\nonumber\\
&=& -H(\mathbf{z})+h_0+\frac{1}{2}w_0,
\end{eqnarray}
where $w_0=\sum_{\bar i<\bar j}w_{\bar i \bar j}$ is total edge weight of $G^*$. In particular, the lowest-energy state of the Ising model, namely the ground state, corresponds to the maximum cut in $G^*$ \cite{Barahona-1988}.

\subsection{Some Algorithms}

Although we could not directly model Max-Cut through linear programming, we could make linear relaxation after quadratic modeling. This leads to the famous Goemans-Williamson (GW) algorithm~\cite{GW-1995}, which is by now the best algorithm for Max-Cut theoretically. For graphs with positive weights, GW algorithms achieves an approximation ratio at least $0.878$. However, the algorithm has a time complexity at least $O(n^3)$, and thus is not efficient enough for practical uses. A truncated version of GW, called CirCut, is proposed in \cite{CirCut}, which contains non-convex optimization steps but performs reasonably well practically.

Alternatively, people have attempted to construct approximate algorithms employing the greedy strategy since the 1970s. Recently we classify these greedy heuristics into two different classes, by making an analogy with the Minimum Spanning Tree problem. Those vertex-oriented algorithms are clarified as Prim class, which contains the earliest Sahni-Gonzalez~(SG) algorithm \cite{SG} and some refined versions, especially SG3~\cite{EC}. Those edge-oriented algorithms are clarified as Kruskal class, which contains mainly the Edge Contraction~(EC) algorithm~\cite{EC} and its further refinements, Differencing Edge Contraction (DEC)~\cite{DEC}, and Signed Edge Contraction (SEC) \cite{SEC,MC-Tree}. In general, these greedy heuristics could be very efficient, with a time complexity $O(n^2)$ or even linear to the total edge number.

With the Ising formulation, it also becomes natural to construct heuristic algorithms for Max-Cut with simulated annealing~(SA)~\cite{SA}. SA could provide very nice numerical results in practice, partially due to its global nature. However, the time complexity and theoretical performance guarantee are not clear at all. In the later calculations with SA, we use the open-source software D-Wave-Neal, version 0.6.0 \cite{D-Wave-neal}.

\subsection{Sherrington-Kirkpatrick model}

We use a specific instance, the SK model, to assess the performance of these algorithm except GW. Roughly speaking, this is an Ising model with random couplings on a complete graph. Explicitly, one sets $h_0=0$ and all $h_i=0$ in (\ref{eq.Ising}), and chooses each coupling $J_{ij}$ independently from an identical distribution satisfying
\begin{equation}
\overline {J_{ij}}=0,\quad \overline{J^2_{ij}}=1.
\end{equation}
In the thermodynamic limit $n\to \infty$, it is well known that the normalized groundstate energy approaches the Parisi value ~\cite{Parisi-1979}
\begin{equation}
\lim_{n\to \infty} \frac{\overline E}{n^{3/2}}=\Pi^*=-0.763166...
\end{equation}
Here the average over different samplings of the couplings has been taken. Unfortunately, Parisi's approach only gives the energy, but not the explicit groundstate.

How could the Parisi value be approximated algorithmically? If one applies the naive greedy heuristic, essentially the original SG algorithm \cite{SG}, one would get the limiting value \cite{Aizenman-1987}
\begin{equation}
\Pi_{\rm{SG}}=-\frac{2}{3}\sqrt{\frac{2}{\pi}}\approx -0.532.
\end{equation}
Or one may apply a rough semi-definite programming (SDP) analysis, which leads to~\cite{Aizenman-1987}:
\begin{equation}
\Pi_{\rm{SDP}}=-\frac{2}{\pi}\approx -0.637.
\end{equation}
Both of them are still far away from the true energy. However, the situation changes dramatically if we adopt the refined algorithms reviewed previously. The results from various algorithms are plotted in the following figure.

\begin{figure}[H]
\centering
	\includegraphics[width=0.85\textwidth]{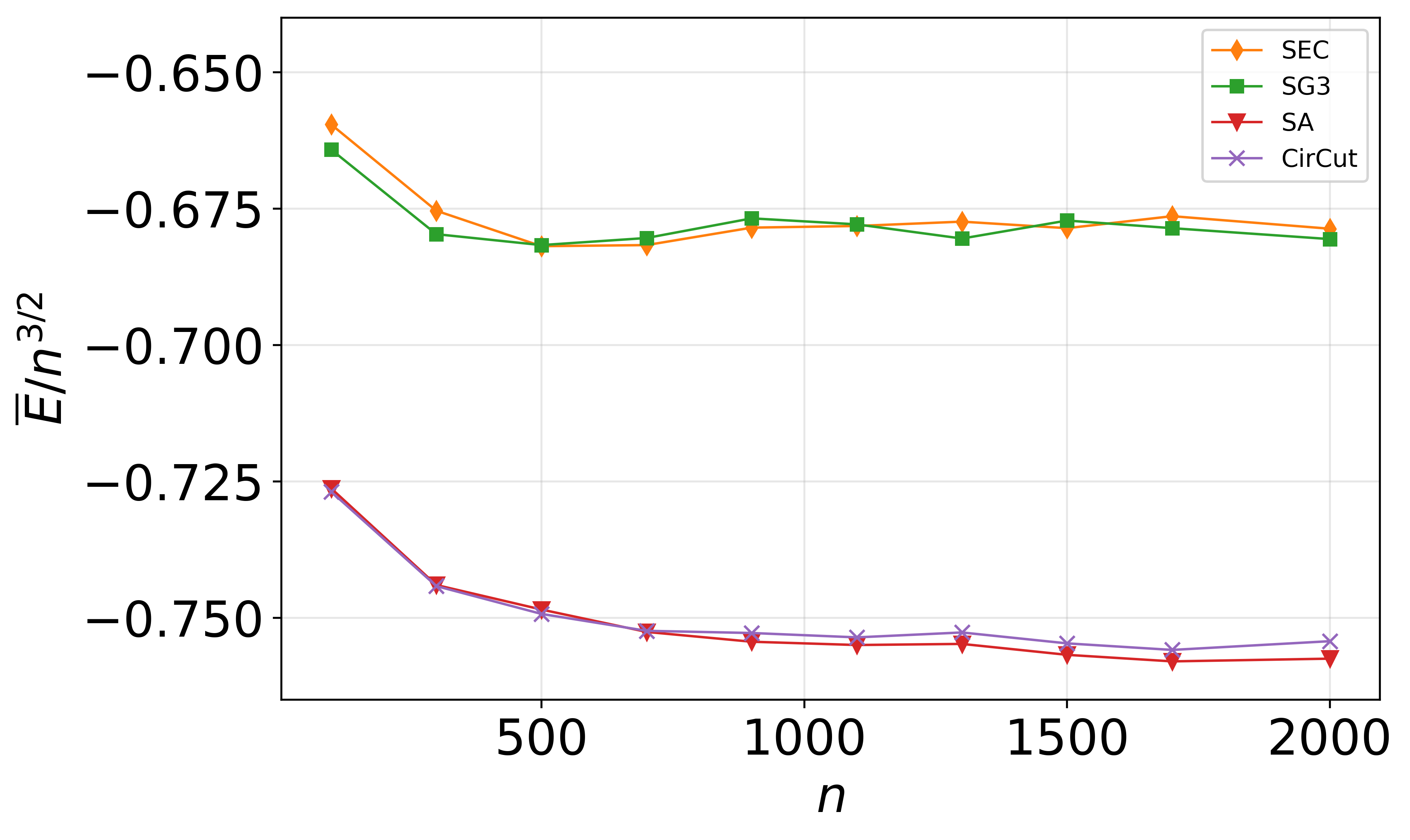}
\caption{\it Average performance of various algorithms for the groundstate energy of the SK model. For each $n$, we generate $20$ instances by sampling $\{w_{ij}\}$ independently from the normal distribution $\mathcal{N}(0,1)$, calculate the normalized energy $\bar{E}/n^{3/2}$ for each instance, and finally take the average over all instances. For SA the number of sweeps $n_s$ has been chosen to be $300000$.}\label{fig:SK}
\end{figure}

From Figure \ref{fig:SK} we see that both SEC and SG3 could achieve a normalized energy as low as $-0.68$, greatly improving the naive SG/greedy value and also the SDP value. CirCut and SA perform even better, approaching energy values of $-0.755$ and $-0.758$ respectively. These results are rather close to the Parisi value. Promoting SA to digital devices, one could further improve its performance, see~\cite{Annealing-2022}.

\section{Maximum Independent Set}

 Theoretically, MIS is strictly equivalent to the Maximum Clique problem, and also equivalent to the Minimum Vertex Cover problem. Here we will focus on MIS specifically. Similar to Max-Cut, it is also not easy to characterise the MIS problem through linear programming~\cite{Matching-1986,BBPP-1999}.

\subsection{Ising modeling}

As shown in the previous section, the Max-Cut problem could be exactly expressed as an QUBO problem, or an Ising model. This is actually also the case for MIS \cite{Barahona-1982,BBPP-1999}. The logic is as follows. In MIS we have edge constraints, which state that no vertices of the same edge should be included in the allowed set. We soften these hard constraints and model them with penalty terms in the cost function or Hamiltonian. Thus, to describe the MIS problem in a graph $G=(V,E)$, we could use the following Hamiltonian of the QUBO form
\begin{equation}
H(\mathbf{x})=\beta\sum_{ij\in E} x_i x_j   - \mu \sum_{i\in V} x_i.
\end{equation}
Following the physical convention, we call $\beta\equiv 1/T$ the inverse temperature, and $\mu$ the chemical potential. As far as we known, such an description of MIS first made its appearance in \cite{Barahona-1982}, where its relation to Ising model is discussed and analyzed.


To see what such a Hamiltonian really counts, we may express it in a different way. Noticing that those vertices with $x_i=0$ do not really contribute to the Hamiltonian at all, we may simply throw them away. Then we are left with
the induced subgraph from the vertices with $x_i=1$. Denote the induced subgraph as $G_1=(V_1,E_1)$, the Hamiltonian reads
\begin{equation}
H(\mathbf{x})=\beta\sum_{ij\in E_1} x_i x_j -\mu \sum_{i\in V_1} x_i =\beta |E_1|-\mu |V_1|.
\end{equation}
Such a function form has been used in early investigations of MIS through SA \cite{Aarts-1989,DIMACS-1996}.

\subsubsection{Parameter Choice}

With the above QUBO formulation, we could already solve it with various algorithms reviewed in the previous section. Before doing so, we would like to discuss the choices of the parameters in more detail.

Two intuitive approaches for dealing with such a Hamiltonian are adopted recently in \cite{Angelini-2019}. In one approach, one takes the zero temperature limit to eliminate the first term, and thus restores the original hard constraints. In the other, one fixes in advance the particle/vertex number to eliminate the second term, and then solve the remaining constrained system to verify the answer. Alternatively, one may take some moderate choices for $\beta$ and $\mu$, or effectively for the ratio $\lambda\equiv\beta/\mu$. In \cite{Lucas-2014} it is suggested that one should always take $\lambda>1$ so that no violation of the constraints would be favored. Indeed, choosing $\lambda=2$ would be strong enough to guarantee that the relaxed problem is equivalent to the original MIS \cite{BBPP-1999}. This thus becomes the popular choice recently \cite{Chapuis-2019,Pelofske-2019}.

However, here we would like to make the critical choice $\lambda=1$, to further relaxing the constraints.
Actually even with such a choice, the relaxed problem would still be exactly equivalent to the original MIS \cite{Barahona-1982}. The underlying idea is that, if with such a choice the energy of the solution is $E_0<0$, one could always achieve an independent set of order $|E_0|$ \cite{Barahona-1982}. Operationally, we could achieve this by pruning the remaining edges in the solution one by one \cite{Wybo-2024}. Some refined scenarios will be discussed later. In fact, in early application of SA to MIS \cite{Aarts-1989}, varying $\lambda$ in the range $[0.7, 1.2]$ is standard \cite{DIMACS-1996}. The critical choice $\lambda=1$ has also been adopted recently in \cite{Zeng-2024,Wybo-2024}.

With the choice $\lambda=1$, the Hamiltonian could be normalized as \cite{Barahona-1982}
\begin{equation}
H(\mathbf{x})=\sum_{ij\in E} x_i x_j   - \sum_{i\in V} x_i. \label{eq.MIS-QUBO}
\end{equation}
We would like to transform it into the Ising form with (\ref{eq.QUBO-Ising}). For this, we set $|V|=n$, $|E|=m$, and label the degree of vertex $i$ as $d_i$. The Ising Hamiltonian could then be expressed as
\begin{equation}
H(\mathbf{z})=\frac{1}{4}\sum_{ij\in E} z_iz_j-\frac{1}{4}\sum_{i\in V}(d_i-2)z_i +\frac{1}{4}(m-2n).\label{eq.MIS-Ising}
\end{equation}
A reduced formula for regular graphs appears in \cite{Wybo-2024}.

\subsection{Greedy Heuristics}

In an Ising Hamiltonian like (\ref{eq.MIS-Ising}), those terms with stronger couplings will be more important. Therefore, those vertices with large degree $d_i$ will be given more focus in such a framework. In particular, if one take a greedy heuristic, such as SG3 or SEC mentioned before, one would prefer to select the vertex with the largest degree at each step. In fact, one could devise a specific greedy heuristic for MIS in this way, which appears long ago \cite{Johnson-1973} and called ``MAX" \cite{Griggs-1983} for short. MAX deletes the vertex with the maximum degree at each step, until no edge exists at all.

MAX could be considered as a ``worst-out" procedure. Instead, we could also adopt a ``best-in" scenario \cite{Johnson-1973}: at each step include the vertex with minimum degree into the object set, and delete all its neighbors. This is called ``MIN'' in
\cite{Griggs-1983}. It turns out that MIN performs rather well for MIS, as analyzed in \cite{HR-1997} with the logan ``Greed is Good". Nevertheless, it is noticed that MAX may behave much worse than MIN \cite{HR-1997}.

\subsubsection{Dense Random graphs}

To assess the performance of the above greedy heuristics, we apply them to the ER graphs $G(n,p=0.5)$.
The independence number of such graphs is proved to be concentrated when $n$ is large \cite{Matula-1976}:
\begin{equation}
\alpha(G(n,p=0.5))\sim 2\log_2 n,
\end{equation}
However, while a randomized procedure naturally produces a value around $\log_2 n$ \cite{Karp-1976}, known algorithms could hardly guarantee a solution strictly greater than that asymptotically \cite{Coja-2015}. Some numerical study on this subject is conducted recently in \cite{Marino-2023}, to check whether we could approach the independence number when $n$ is not quite large. In this case we should take the finite $n$ corrections into account. As a result, we obtain an extended expansion $R(n)$ for the independence number as \cite{Matula-1976}
 \begin{equation}
 R(n)=2\log_2 n-2\log_2\log_2 n +2\log_2 e -1.
 \end{equation}

\begin{figure}[H]
\centering
	\includegraphics[width=0.85\textwidth]{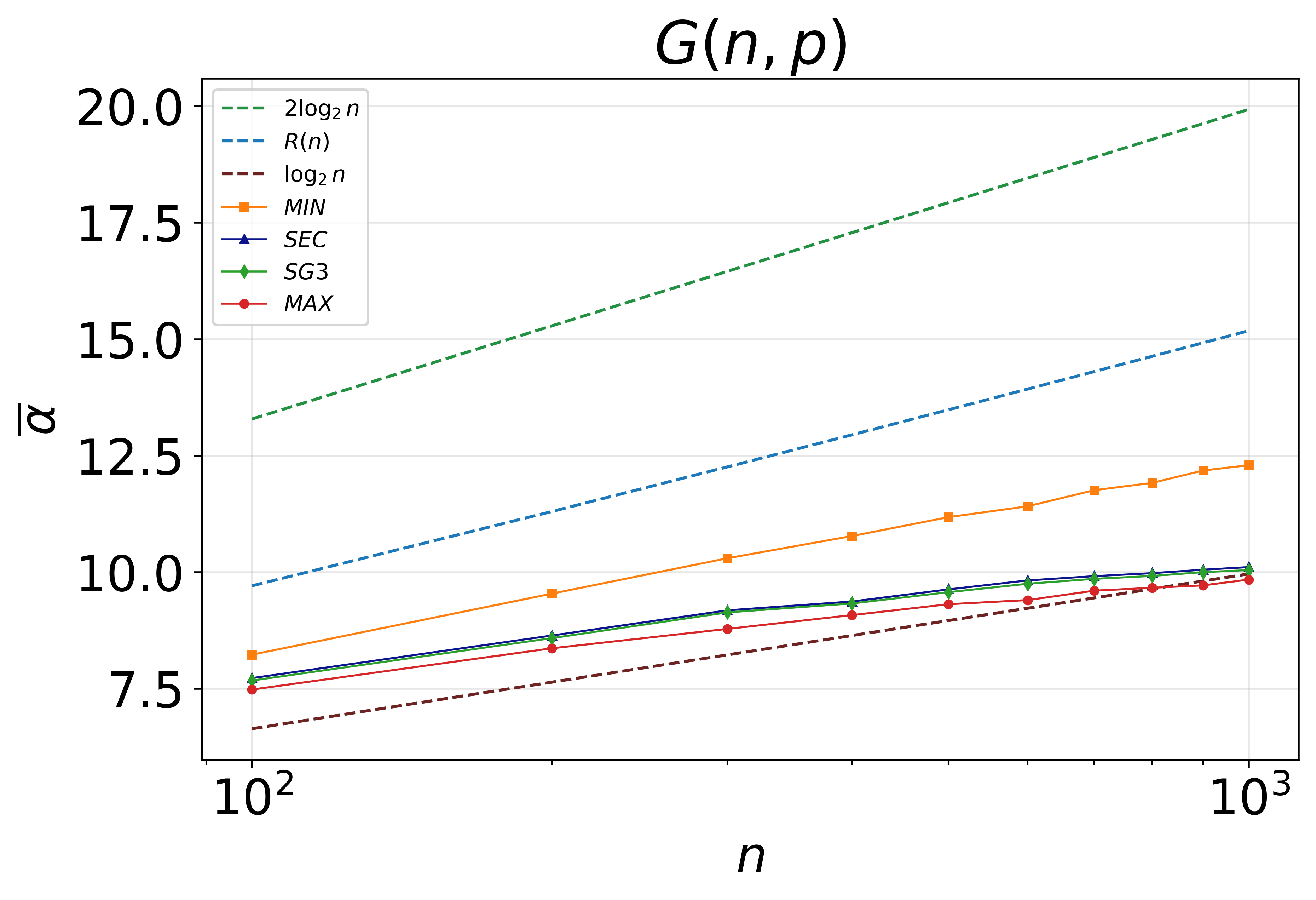}
\caption{\it Average performance of various greedy heuristics for MIS of the ER graphs $G(n,p=0.5)$. For each $n$ we generate $500$ instances, calculate the approximate independence number for each instance, and take the average over instances in the end.}\label{fig:ER-dense}
\end{figure}

 Here we would like to use this problem to assess the performance of different heuristics. As in \cite{Marino-2023}, we generate the ER graphs $G(n,p=0.5)$ in the range $100\le n\le 1000$, and then solve them with various greedy heuristics. When applying the Ising/Max-Cut solvers such as SG3 and SEC, we add a pruning process to the solutions obtained as in \cite{Wybo-2024}. Actually we use MIN or MAX as a filter to refine the solution into an independent set. This would slightly improve the performance compared to the naive pruning procedure used in \cite{Wybo-2024}. The results are plotted in Figure \ref{fig:ER-dense}.

As shown in Figure \ref{fig:ER-dense}, the curve from $R(n)$ lie roughly in the middle of $2\log_2 n$ and $\log_2 n$, demonstrating that the sub-leading terms are significant in this region of $n$ \cite{Marino-2023}. The numerical results in Figure \ref{fig:ER-dense} confirms the assertion in \cite{HR-1997} that MIN performs much better than MAX. However, the MIN curve are far from the theoretical prediction $R(n)$. The results also confirm our speculation that in the present framework, SG3 and SEC would follow a similar procedure as MAX. Indeed, the curves produced by these three heuristics almost coincide with each other,just as expected. Besides, all these three curves quickly drop to the $\log_2 n$ line, which means the ``MAX'' choice hardly provide any advantage. Meanwhile, the MIN curve does not fall down at all, indicating that it may achieve a multiplicative improvement over $\log_2 n$ asymptotically. Similar behavior was observed in \cite{Marino-2023}.

\subsection{More general algorithms}

The results in Figure \ref{fig:ER-dense} seems to indicate that the Ising formulation does not provide any advantage. If we could not even outperform the simplest greedy heuristic, MIN, then the whole framework would be useless. Fortunately, when we go beyond the greedy heuristic, the approximation for MIS could be improved quite a lot, as demonstrated recently in \cite{Angelini-2019}. Encouraged by \cite{Angelini-2019}, we now try to solve MIS with the other Max-Cut solvers, such as SA and CirCut.

\subsubsection{Dense random graphs}

We still use the previous random ER graphs $G(n,p=0.5)$ as a starting point. So we now repeat the whole calculation with SA, and check if it outperforms MIN or not. We notice that similar investigations have actually been performed in \cite{DIMACS-1996}, though without explicit use of an Ising Hamiltonian. Besides, in \cite{DIMACS-1996} SA has been restricted to take roughly the same amount of time as MIN. As a result, the improvement over MIN is not quite significant. Here we relax this constraint, and vary the critical parameter $n_s$, number of sweeps, in the range $[1000, 300000]$. This makes the running time of SA about $5-500$ times longer than MIN. The results for $n_s=1000$ and $n_s=300000$ are plotted in the following figure, together with those from MIN.

\begin{figure}[h]
\centering
	\includegraphics[width=0.85\textwidth]{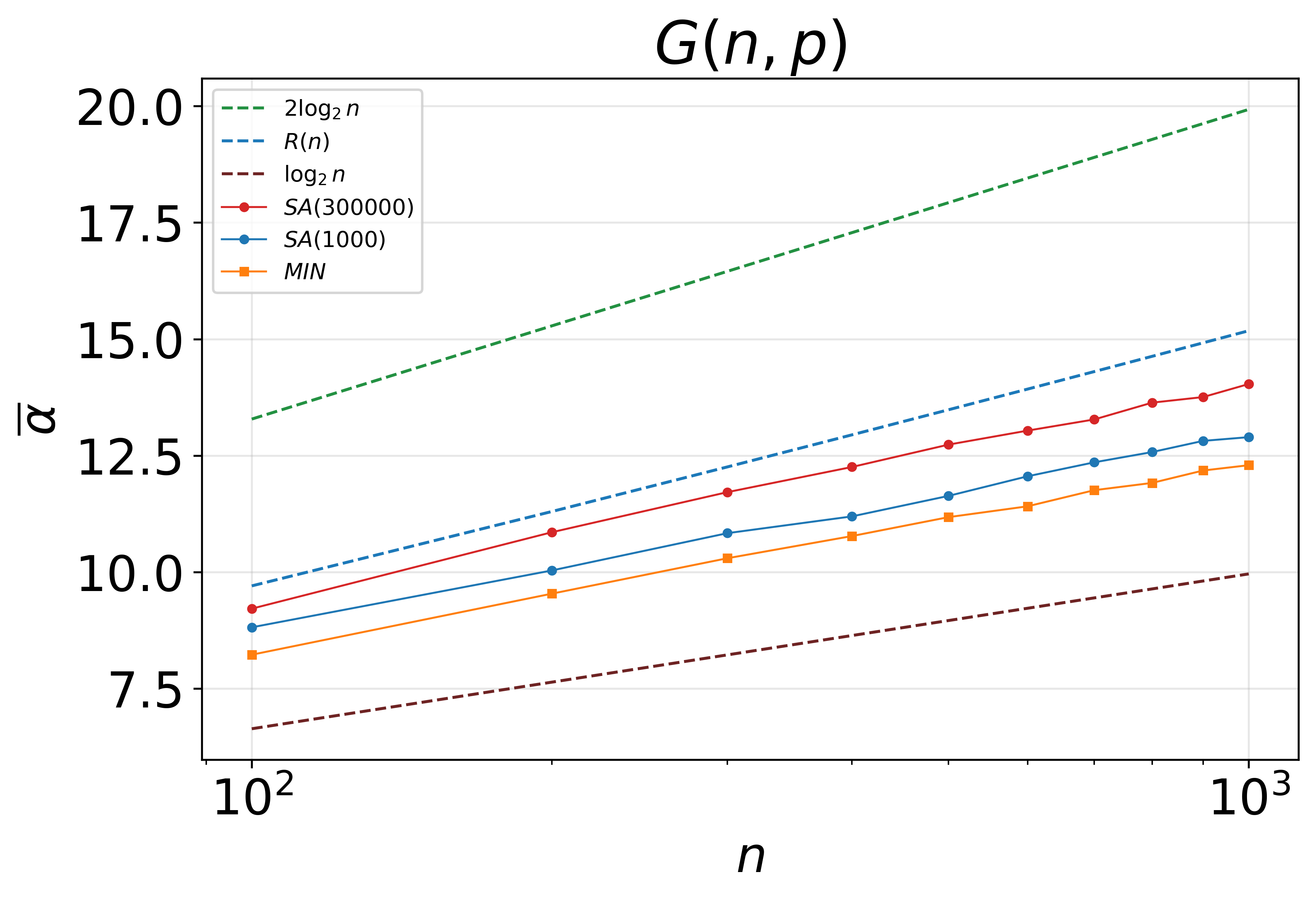}
\caption{\it Average performance of SA for MIS of the ER graphs $G(n,p=0.5)$, in comparison with that of MIN. Here we generate $50$ instances for each $n$, calculate the approximate independence number for each instance, and take the average over the instances in the end. The curve of SA with $n_s=1000$ is labeled as SA(1000), and that with $n_s=300000$ is labeled as SA(3000000). }\label{fig:ER-dense2}
\end{figure}

Quite different from the MAX, SG3 and SEC curves in Figure \ref{fig:ER-dense}, the above figure shows that SA could easily surpass the results from MIN, and could be gradually improved by increasing $n_s$. This gives us confidence that the present Ising framework does work, even though the greedy heuristic does not fit in anymore.

Still one may worry about the much longer time that SA takes. We believe that such an overload is rather modest, since it could be taken to be independent of the input size. It is recently found in \cite{Marino-2023} that the results from MIN could be greatly improved by starting from a randomly-chosen one-vertex or two-vertex set. However, the time complexity would be increased by a factor of $n$ or $n^2$, respectively. Compared to those scenarios, the running time of SA would be quite acceptable.

\subsubsection{Sparse random graphs}


 We will further study sparse ER graphs $G(n,p)$ with much lower average degree $\bar d=np$. When $1\ll \bar d \ll n$, it has been proved that the independence number concentrates as \cite{Frieze-1990}:
\begin{equation}
\alpha(G(n,\bar d/n))\sim \frac{2n}{\bar d}(\log \bar d -\log\log \bar d -\log 2+1).\label{eq.alpha-asymptotic}
\end{equation}
 Therefore in such a region, the independence number is almost linear in $n$, and exhibits complicated behavior with $\bar d$. Due to this, we may define the independence density (or independence ratio as used by mathematicians) $\rho\equiv \alpha/n$, and focus on its $\bar d$ dependence. When $\bar d$ is large enough, the above expansion is dominated by the leading term
\begin{equation}
\alpha(G(n,\bar d/n))\sim \frac{2n}{\bar d}\log \bar d .\label{eq.alpha-asymptotic2}
\end{equation}
However, no known algorithm could guarantee a solution whose size would be larger than one
half of this asymptotically \cite{Coja-2015}. Since the logarithmic increasing is rather slow, we may instead focus on the intermediate region where $\bar d$ is not very large. Then the problem turns into this: could any algorithm provide approximate solutions close to the full expansion (\ref{eq.alpha-asymptotic}) numerically? The recent investigation in \cite{Angelini-2019} indicates that the answer would be positive, especially when Monte Carlo algorithms are employed. So we solve MIS in this set of graphs with refined Ising solvers, specifically SA, and plot the results in the following figure. The results from MIN are also plotted for comparison.

\begin{figure}[h]
\centering
	\includegraphics[width=0.85\textwidth]{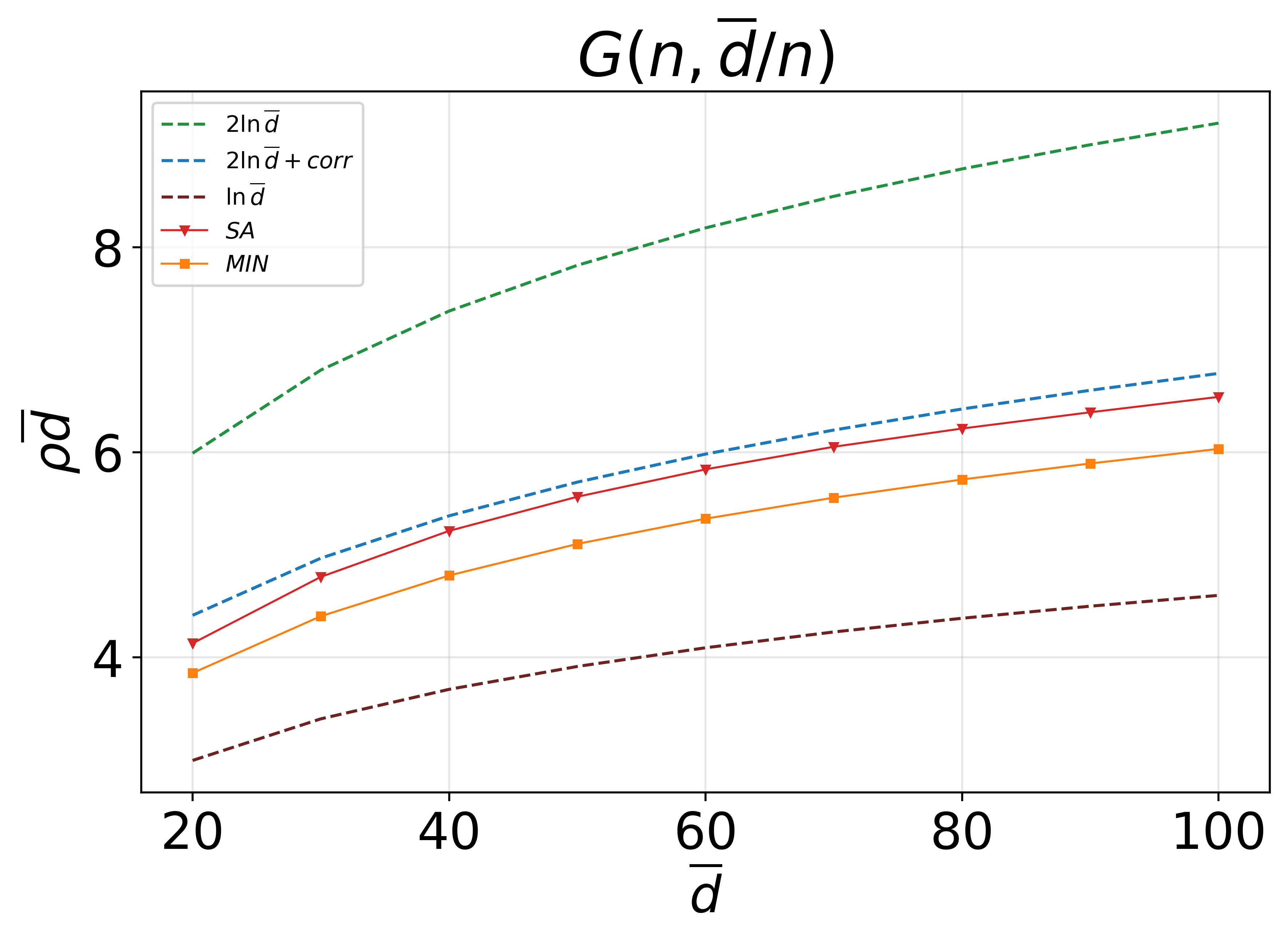}
\caption{\it Average performance of SA, as a specific Max-Cut/Ising solver, for MIS of the sparse ER graphs $G(n,p=\bar d/n)$. Here the number of sweeps in SA is chosen as $n_s=1000$. For each value of $\bar d$, we fix the graph size to be $n={\bar d}^2$. Then for each pair $(n,\bar d)$ we generate $500$ instances, calculate the independence density, and take the average over instances in the end. The predictions from the formulae $2\ln \bar d$, $\ln \bar d$ and the full expansion (\ref{eq.alpha-asymptotic}) (denoted as $2\ln \bar d +corr$) are plotted as dashed lines for comparison.
 }\label{fig:ER-sparse}
\end{figure}

Figure \ref{fig:ER-sparse} resembles Figure \ref{fig:ER-dense2} in several aspects. First, for the parameter region considered, namely $20\le \bar d \le 100$ and $n=\bar d ^2$, the curve from the full expansion (\ref{eq.alpha-asymptotic}) is much lower than that from (\ref{eq.alpha-asymptotic2}). This again demonstrates that in such region the sub-leading terms make non-negligible contributions. Secondly, the prediction from MIN is already not far from the theoretical curve, confirming again the conclusion made in \cite{HR-1997}. Most importantly, the results from SA are very close to the theoretical curve, even though the number of sweeps $n_s$ is as low as $1000$. With larger values of $n_s$, one would certainly achieve further improvement. All these results are in accordance with those in \cite{Angelini-2019}, though the underlying graphs are not quite the same.

\subsubsection{Random Regular Graphs}
Now we turn to a special class of graphs, the random regular graphs $G_d$. MIS for such a class of graphs has been intensively studied \cite{Barbier-2013}\cite{Angelini-2019}. Since regular graphs constitute a special subset of those ER graphs, one may wonder how the independence number of $G_d$ behaves when $d$ is large. In fact, in the regular case we could do much better and obtain the exact independence number at the level of physical rigor \cite{Barbier-2013}. The derivation uses a refined strategy of the original Parisi procedure \cite{Parisi-1979}, which is called one step replica symmetry breaking (1RSB) \cite{Barbier-2013}. The 1RSB results are determined by a set of complicated equations, which could be solved numerically. For example, taking $n=10^5$, the independence density  $\rho_{\textrm{1RSB}}$ reads \cite{Barbier-2013}:

\begin{table}[h]
\begin{center}
\begin{tabular}{||c||c|c|c|c|c|c|c|c|c||}
\hline
  $d$ &  $20$ &   $30$  &  $40$    &  $50$  & $60$  &  $70$  &  $80$   &   $90$  &   $100$  \\
\hline
    $\rho_{\textrm{1RSB}}$     &  $0.1948$  &  $0.1529$   &  $0.1273$ &  $0.1098$  & $0.0970$  &  $0.0871$ &  $0.0792$  &  $0.0728$  &  $ 0.0674$  \\
\hline
\end{tabular}
\caption{The 1RSB prediction for the independence density in random regular graphs $G_d$.}\label{Tab:1RSB}
\end{center}
\end{table}

Moreover, one could derive the asymptotic expansion for $\alpha_{\textrm{1RSB}}(G_d)$ at large $d$. The resulting expansion turns out to coincide with (\ref{eq.alpha-asymptotic}) \cite{Barbier-2013}. This seems to indicate that $\alpha_{\textrm{1RSB}}$ may represent the exact results for MIS in the general case of random ER graphs. To check this, we insert the data in Table \ref{Tab:1RSB} into Figure \ref{fig:ER-sparse}, and obtain the following figure:

\begin{figure}[H]
\centering
	\includegraphics[width=0.85\textwidth]{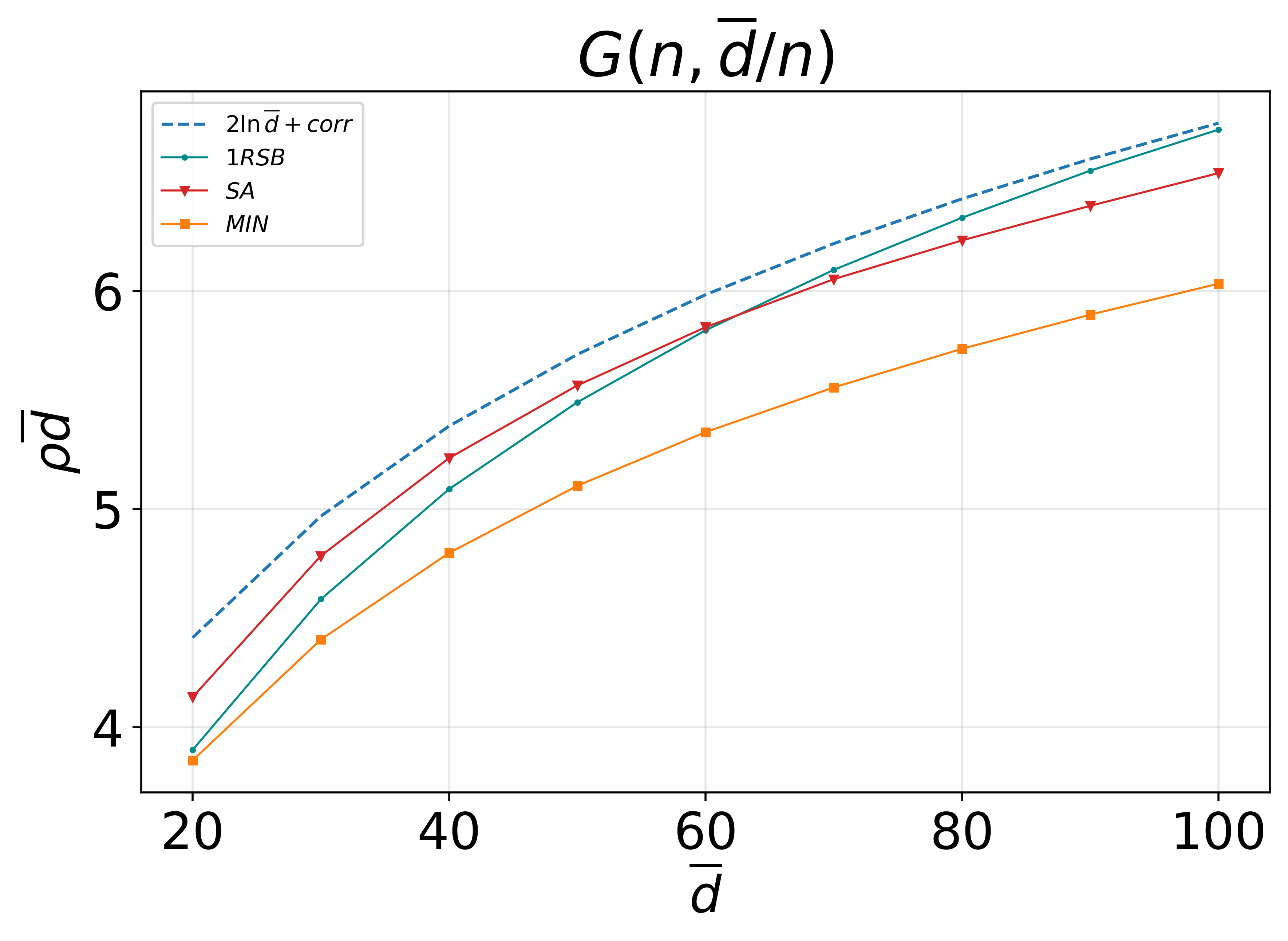}
\caption{\it Direct comparison of the results obtained for the ER graphs $G(n,p=\bar d/n)$ in Figure \ref{fig:ER-sparse}, and the 1RSB prediction for the random regular graphs $G_d$ in Table \ref{Tab:1RSB}.}\label{fig:1RSB}
\end{figure}

The above figure shows clearly that, while the MIN curve is always below the 1RSB prediction, the results from SA surpass the 1RSB prediction in the range $20\le \bar d \le 60$. Since the 1RSB prediction is believed to be exact when $d\ge20$ \cite{Barbier-2013,Ding-2013}, this implies that the independence density of the random ER graphs must be larger than the random regular ones when $20\le \bar d \le 60$. Therefore, the 1RSB prediction could not be naively applied to the random ER graphs, as we intuitively guessed before. In order to extend the 1RSB procedure to the ER graphs, at least some modifications or refinements would be needed.

\newpage
\subsubsection{Special Benchmarks}

Now we turn to deterministic instances. In a recent review article on MIS \cite{Marino-2024}, several popular benchmarks used in the literature are provided. Among those benchmarks, the one arising from coding theory \cite{Sloane-2000} is considered to be very challenging. We now use this benchmark to further assess the performance of those algorithms discussed previously. The detailed information of the instances in this benchmark is list in Table \ref{Tab:code-input}. Notice that the independence number of 1dc.2048 has recently been confirmed to be $172$ 
\cite{No-2019,Nakasho-2023}, and the upper bound for 1dc.4096 has been lowered to $320$~\cite{No-2019}.

\begin{table}[h]
\centering
\begin{minipage}{.5\linewidth}
\centering
\begin{tabular}{ |l|l|l|l| }
\hline
Name & $|V|$ & $|E|$ & $\alpha(G)$ \\
\hline
1dc.64 & 64 &  543 & 10 \\
1dc.128 & 128 &  1471 & 16 \\
1dc.256 & 256 &  3839 & 30 \\
1dc.512 & 512 &  9727 & 52 \\
1dc.1024 & 1024 &  24063 & 94 \\
1dc.2048 & 2048 &  58367 & 172 \\
1dc.4096 & 4096 & 139263  & 316-320\\
2dc.128 & 128 &  5173 & 5 \\
2dc.256 & 256 &  17183 & 7 \\
2dc.512 & 512 &  54895 & 11 \\
2dc.1024 & 1024 &  169162 & 16 \\
2dc.2048 & 2048 &  504451 & 24 \\
1tc.8 & 8 &  6 & 4 \\
1tc.16 & 16 &  22 & 8 \\
1tc.32 & 32 & 68 & 12 \\
1tc.64 & 64 &  192 & 20 \\
1tc.128 & 128 &  512 & 38 \\
\hline
\end{tabular}
\end{minipage}%
\begin{minipage}{.5\linewidth}
\centering
\begin{tabular}{ |l|l|l|l| }
\hline
Name & $|V|$ & $|E|$ & $\alpha(G)$ \\
\hline
1tc.256 & 256  &  1312 & 63 \\
1tc.512 & 512 &  3264 & 110 \\
1tc.1024 & 1024 & 7936 & 196 \\
1tc.2048 & 2048 & 18944 & 352 \\
1et.64 & 64 & 264 &18 \\
1et.128 & 128 & 672 &28 \\
1et.256 & 256 & 1664 & 50 \\
1et.512 & 512 & 4032 & 100 \\
1et.1024 & 1024  & 9600 & 171 \\
1et.2048 & 2048 & 22528 & 316 \\
1zc.128 & 128 & 2240 & 18 \\
1zc.256 & 256 & 5632 & 36 \\
1zc.512 & 512 & 13824 & 62 \\
1zc.1024 & 1024 & 33280 & 112--117 \\
1zc.2048 & 2048 & 78848 & 198--210 \\
1zc.4096 & 4096 & 184320 & 379--410\\
\hline
\end{tabular}
\end{minipage}
\caption{Graphs arising from coding theory.}\label{Tab:code-input}
\end{table}
\begin{table}[h]
\centering
\begin{minipage}{.5\linewidth}
\centering
\begin{tabular}{ |l|l|l|l|l| }
\hline
Name & MIN & CirCut & SA   & $\alpha(G)$ \\
\hline
1dc.64 & 10 &  10 & 10 & 10 \\
1dc.128 & 15 &  16 & 16 & 16 \\
1dc.256 & 26 &  30 & 30 & 30 \\
1dc.512 & 43 &  52 & 52 & 52 \\
1dc.1024 & 77 &  93 & 94  & 94 \\
1dc.2048 & 131 &  171 & 172  & 172 \\
1dc.4096 & 236 &  315 & 316  & 316-320 \\
2dc.128 & 5 &  5 &5  & 5 \\
2dc.256 & 7 &  7 &7 & 7 \\
2dc.512 & 10 &  10 & 11 & 11 \\
2dc.1024 & 15 &  14 & 16 & 16 \\
2dc.2048 & 21 &  21 & 24 & 24 \\
1tc.8 & 4 &  4 & 4 & 4 \\
1tc.16 & 8 &  8 & 8  & 8 \\
1tc.32 & 12 & 12 & 12  & 12 \\
1tc.64 & 20 &  20 & 20  & 20 \\
1tc.128 & 38 &  38 & 38  & 38 \\
\hline
\end{tabular}
\end{minipage}%
\begin{minipage}{.5\linewidth}
\centering
\begin{tabular}{ |l|l|l|l|l| }
\hline
Name & MIN & CirCut &  SA   &  $\alpha(G)$ \\
\hline
1tc.256 & 61  &  62 & 63  & 63 \\
1tc.512 & 106 &  110 & 110  & 110 \\
1tc.1024 & 189 & 187 & 196  & 196 \\
1tc.2048 & 331 & 331 & 352  & 352 \\
1et.64 & 18 & 18 & 18  &18 \\
1et.128 & 28 & 28  & 28  &28 \\
1et.256 & 50 & 50 &  50  & 50 \\
1et.512 & 96 & 98 & 100  & 100 \\
1et.1024 & 155  & 167 & 171  & 171 \\
1et.2048 & 293 & 296 & 316  & 316 \\
1zc.128 & 16 & 18 & 18 & 18 \\
1zc.256 & 36 & 36 & 36  & 36 \\
1zc.512 & 58 & 62 & 62  & 62 \\
1zc.1024 & 103 & 108 & 112  & 112--117 \\
1zc.2048 & 181 & 178 &  198  & 198--210 \\
1zc.4096 & 329 & 322 & 379   & 379--410\\
\hline
\end{tabular}
\end{minipage}
\caption{Computational results for graphs arising from coding theory. The results shown are the largest value in 50 runs for each algorithm.}\label{Tab:code-output}
\end{table}

We solve these instances with three algorithms: MIN, CirCut, and SA.  Since all of them contain randomized steps, we
 run them $50$ times for each instance, and take the largest value as the output. We also set the number of sweeps in SA to $n_s=400000$, in order to obtain the optimal value. The results are summarized in Table \ref{Tab:code-output}.

From the results in Table \ref{Tab:code-output} one can see that the performance of the three algorithms differ a lot. On average, CirCut performs better than MIN, especially for these four cases: 1dc.512, 1dc.1024, 1dc.2048 and 1dc.4096 (Here 1dc refers to Single-Deletion-Correcting Codes \cite{Sloane-2000,Sloane-2002}). The performance of SA is even better than CirCut.
Specifically, SA obtains the optimal values for all the instances whose independence number is known. And for the remaining four graphs, namely 1dc.4096, 1zc.1024, 1zc.2048 and 1zc.4096, the predictions from SA coincide with the lower bounds known. In essence, the best results from SA are exactly the same as those from a local search strategy long ago \cite{Butenko-2009}~(except for 1dc.4096 which is not calculated in \cite{Butenko-2009}). This gives more confidence that these lower bounds could actually be optimal.

Furthermore, SA could provide additional information for the corresponding codes. Let us check the results for the 1dc graphs more carefully. It is known \cite{Sloane-2000,Sloane-2002} that, for $n\le 10$ the Varshamov-Tenegolts code \cite{VT-1965} $\mbox{VT}_0(n)$ are optimal for these graphs. However, they are not the unique solutions. For example, one may replace the codes $(110100)_2$ and $(001011)_2$ in $\mbox{VT}_0(6)$ by $(111000)_2$ and $(000111)_2$, and obtain another MIS \cite{Sloane-2002}. Our solutions suggest that we can do the same for $\mbox{VT}_0(10)$, by replacing the corresponding decimal codewords $95$ and $928$ with $63$ and $960$. Moreover, we now know $\mbox{VT}_0(11)$ gives a MIS of 1dc.2048 \cite{Sloane-2000, No-2019,Nakasho-2023}. Our calculation produces another independent set of size $172$, and it contains almost no common vertices with $\mbox{VT}_0(11)$.

Unfortunately, with the present choice of parameters, especially the number of sweeps, the running of SA is slowed down. For the graph with the largest size, 1zc.4096, it takes about 4500s to finish a single run of SA on a desktop computer with a 8-core CPU and 64 GB RAM. This would be much longer than the time taken in the local search strategy \cite{Butenko-2002}\cite{Butenko-2009}. Properly designing the annealing schedule could possibly reduce the running time, but this is beyond the scope of the present paper.

\section{Summary}

The close relation between the MIS problem and the Ising model allows us to investigate MIS with various Ising/Max-Cut solvers.  With the solvers elaborately chosen, we greatly improve the approximation for the independence number in various graphs. For both the dense ER graphs $G(n,p=0.5)$ and sparse ER graphs $G(n,p=\bar d/n)$ with $\bar d \ll n$, we show that the independence number could be well approximated numerically, although not asymptotically. And for a difficult benchmark arising from coding theory, the Ising/Max-Cut solvers also exhibit remarkable performance, especially SA.

The true advantage of the present Ising formalism would appear when the weights are introduced. This would include the weighted version of MIS, and both the vertex-weighted and edge-weighted versions of the Maximum Clique problem. All these problems could be naturally accommodated within the present framework. It would be very interesting to explore the performance of various solvers in these cases.

An even interesting question would be, could we develop analogous quantum algorithms along this line, which would provide further improvement on MIS? The Ising formulation is very convenient for developing quantum solvers, either quantum annealers, or circuit-based algorithms. The remarkable performance of SA shown in this paper suggests that the corresponding quantum algorithms could bring great benefits. Could they help to settle those unsolved problems arising from the study of codes \cite{Sloane-1989}, such as the optimal values of the independence number in Table \ref{Tab:code-input}? We hope the answer would be positive, eventually.

\section*{Note added:}

After finishing the paper, we learn that ref.~\cite{Krpan-2024} has already studied the choice of the penalty parameter $\lambda$ and the post-processing procedure in great detail. The preferred choice $\lambda=1$ in \cite{Krpan-2024} coincides with ours. It would be interesting to further investigate the benefits of the post-processing procedure proposed in \cite{Krpan-2024}.




\begin{thebibliography}{100}
\bibitem{Barahona-1982}
F. Barahona,
\newblock On the computational complexity of Ising spin glass models,
\textit{J. Phys. A: Math. Gen.} 15 (1982) 3241-3523.

\bibitem{Welsh-1993}
D. Welsh, 1993,
\newblock{\em Complexity: Knots, Colourings and Counting},
\newblock Cambridge University Press.


\bibitem{Annealing-2022}
Hiroki Oshiyama, Masayuki Ohzeki, 2022,
\newblock Benchmark of quantum inspired heuristic solvers for quadratic unconstrained binary optimization,
\textit{Scientific Reports} (2022) 12:2146.


\bibitem{GW-1995}
M. Goemans and D.P. Williamson, 1995,
\newblock Improved approximation algorithms for MAX-CUT and satisfiability
problems using semidefinite programming,
\textit{Journal of the ACM}, 42:1115-1145, 1995.

\bibitem{Hastad-1999}
J. H{\aa}stad,
\newblock Clique is hard to approximate within $n^{1-\epsilon}$,
\textit{Acta Math.} 182 (1999) 105-142.

\bibitem{SK-1975}
D. Sherrington and S. Kirkpatrick, 1975,
\newblock Solvable Model of a Spin-Glass,
\textit{Phys. Rev. Lett.} 35 (1975) 1792-1796.


\bibitem{Dantzig-1963}
George B. Dantizig, 1963,
\newblock {\em Linear Programming and Extensions},
\newblock Princeton University Press, Princeton, New Jersey.

\bibitem{Cook-CO}
W. Cook, W. Cunningham, W. Pulleyblank and A. Schrijver, 1998,
\newblock{\em Combinatorial Optimization},
\newblock John Wiley \& Sons, Inc.

\bibitem{MQLib}
I. Dunning, S. Gupta, and J. Silberholz, 2018,
\newblock What Works Best When? A Systematic Evaluation of Heuristics for Max-Cut and QUBO,
\textit{INFORMS journal on computing} 30 (2018) 608-624.




\bibitem{Barahona-1988}
F. Barahona, M. Gr\"{o}tschel, M. J\"{u}nger, and G. Reinelt, 1988,
\newblock An application of connbinatorial
optimization to statistical physics and circuit layout design,
\textit{Oper. Res.} 36, 493-513.


\bibitem{CirCut}
S. Burer, R. Monteiro and Y. Zhang, 2001,
\newblock Rank-two relaxation heuristics for Max-Cut and other binary quadratic programs,
\textit{SIAM J. OPTIM.} 12 (2001) 503-521

\bibitem{SG}
S. Sahni, T. Gonzales, 1976,
\newblock P-complete approximation problems,
\textit{Journal of the ACM}, 23 (1976), 555-565.


\bibitem{EC}
Sera Kahruman, Elif Kolotoglu, Sergiy Butenko, Illya V. Hicks, 2007,
\newblock On greedy construction heuristics for the MAX-CUT problem,
\textit{Int. J. Comput. Sci. Eng.} 3 (2007) 211-218.

\bibitem{DEC}
Refael Hassin, Nikita Leshenko, 2021,
\newblock Greedy Differencing Edge-Contraction heuristic for the Max-Cut
problem,
\textit{Operations Research Letters} 49 (2021) 320-325.


\bibitem{SEC}
Chuixiong Wu, Jianan Wang, Fen Zuo, 2023,
\newblock Stabilizer Approximation III: Maximum Cut,
[\href{https://arxiv.org/abs/2303.17215}{https://arxiv.org/abs/2303.17215}].

\bibitem{MC-Tree}
Jianan Wang, Chuixiong Wu, Fen Zuo, 2023,
\newblock More on greedy construction heuristics for the MAX-CUT problem,
[\href{https://arxiv.org/abs/2312.10895}{https://arxiv.org/abs/2312.10895}].


\bibitem{SA}
S. Kirkpatrick, C. D. Gelatt, Jr., M. P. Vecchi, 1983,
\newblock Optimization by Simulated Annealing,
\textit{Science} 220 (1983) 671-680.


\bibitem{D-Wave-neal}
\newblock dwave-neal,
[\href{https://github.com/dwavesystems/dwave-neal}{https://github.com/dwavesystems/dwave-neal}].



\bibitem{Parisi-1979}
G. Parisi,
\newblock Infinite number of order parameters for spin glasses,
\textit{Phys. Rev. Lett.} 43 (1979) 1754.

\bibitem{Aizenman-1987}
M. Aizenman, J. L. Lebowitz, and D. Ruelle, 1987,
\newblock Some rigorous results on the sherrington-kirkpatrick spin glass model,
\textit{Comm. Math. Phy.} 112 (1987), 3-20.


\bibitem{Matching-1986}
L. Lov\'{a}sz, and M. D. Plummer, 1986,
\newblock {\em Matching Theory},
\newblock Akademiai Kiado, Budapest.


\bibitem{BBPP-1999}
Immanuel M Bomze, Marco Budinich, Panos M Pardalos, and Marcello Pelillo, 1999,
\newblock The maximum clique problem,
\textit{Handbook of Combinatorial Optimization: Supplement Volume A}, pp. 1-74, 1999.

\bibitem{Aarts-1989}
E. Aarts and J. Korst, 1989,
\newblock{\em Simulated Annealing and Boltzmann Machines},
\newblock J. Wiley \& Sons, Chichester, UK, 1989.

\bibitem{DIMACS-1996}
S. Homer and M. Peinado, 1996,
\newblock Experiments with Polynomial-time CLIQUE Approximation Algorithms on Very Large Graphs,
In: \textit{Cliques, coloring, and satisfiability: second DIMACS implementation challenge}, (1996) 147-167.




\bibitem{Angelini-2019}
Maria Chiara Angelini and Federico Ricci-Tersenghi, 2019,
\newblock Monte Carlo algorithms are very effective in finding the largest independent set in sparse random graphs,
\textit{Phys. Rev. } E100 (2019) 013302.


\bibitem{Lucas-2014}
A. Lucas, 2014,
\newblock Ising formulations of many NP problems,
\textit{Frontiers in Physics} 2 (2014) 1-15.


\bibitem{Chapuis-2019}
Guillaume Chapuis, Hristo N. Djidjev, Georg Hahn, Guillaume Rizk, 2019,
\newblock Finding Maximum Cliques on the D-Wave Quantum Annealer,
\textit{J Sign Process Syst} 91 (2019) 363-377.
[\href{https://arxiv.org/abs/1801.08649}{https://arxiv.org/abs/1801.08649}].




\bibitem{Pelofske-2019}
E. Pelofske, G. Hahn and H. Djidjec, 2019,
\newblock Solving Large Maximum Clique Problems on a Quantum Annealer,
In: \textit{Feld, S., Linnhoff-Popien, C. (eds) Quantum Technology and Optimization Problems. QTOP 2019. Lecture Notes in Computer Science}, vol 11413.
[\href{https://arxiv.org/abs/1901.07657}{https://arxiv.org/abs/1901.07657}].

\bibitem{Zeng-2024}
Yuhan Huang, Ferris Prima Nugraha, Siyuan Jin, Yichi Zhang, Bei Zeng, Qiming Shao, 2024,
\newblock Quantum Graph Optimization Algorithm,
[\href{https://arxiv.org/abs/2404.06434}{https://arxiv.org/abs/2404.06434}].


\bibitem{Wybo-2024}
Elisabeth Wybo, Martin Leib, 2024,
\newblock Missing Puzzle Pieces in the Performance Landscape of the Quantum Approximate Optimization Algorithm,
[\href{https://arxiv.org/abs/2406.14618}{https://arxiv.org/abs/2406.14618}].


\bibitem{Johnson-1973}
D. S. Johnson, 1973,
\newblock Approximation algorithms for combinatorial problems,
\textit{J. Comput. System Sci.} 9 (1974), 256-278.

\bibitem{Griggs-1983}
J. Griggs, 1983,
\newblock Lower Bounds on the Independence Number in Terms of the Degrees,
\textit{JOURNAL OF COMBINATORIAL THEORY}, Series B 34 (1983), 22-39.


\bibitem{HR-1997}
M. M. Halld\'{o}rsson and J. Radhakrishnan, 1997,
\newblock Greed is Good: Approximating Independent Sets in
Sparse and Bounded-Degree Graphs,
\textit{Algorithmica} 18 (1997) 145-163.


\bibitem{Matula-1976}
D. W. Matula, 1976,
\newblock  The largest clique size in a random graph,
\textit{Technical Report CS 7608}, Department of Computer Science, Southern Methodist
University, 1976.


\bibitem{Karp-1976}
R. M. Karp, 1976,
\newblock The probabilistic analysis of some combinatorial search algorithms,
\textit{Algorithms Complex. New Dir. Recent Results} 1 (1976) 19.


\bibitem{Coja-2015}
A. Coja-Oghlan and C. Efthymiou, 2015,
\newblock On independent sets in random graphs,
\textit{Random Struct. Algor.} 47 (2015) 436-486.



\bibitem{Marino-2023}
Raffaele Marino and Scott Kirkpatrick, 2023,
\newblock  Hard optimization problems have soft edges,
\textit{Scientific Reports}, 13 (2023): 3671.



\bibitem{Frieze-1990}
A. M. Frieze, 1990,
\newblock On the Independence number of random graphs,
\textit{Discrete Math.} 81 (1990) 171-175.


\bibitem{Barbier-2013}

Jean Barbier, Florent Krzakala, Lenka Zdeborov\'{a} and Pan Zhang, 2013,
\newblock The hard-core model on random graphs revisited,
\textit{J. Phys.: Conf. Ser.} 473 (2013) 012021.


\bibitem{Ding-2013}
Jian Ding, A. Sly, Nike Sun, 2013,
\newblock Maximum independent sets on random regular graphs,
\textit{Acta Mathematica} 217 (2016) 263-340.


\bibitem{Marino-2024}
Raffaele Marino, Lorenzo Buffoni, Bogdan Zavalnij, 2024,
\newblock A Short Review on Novel Approaches for Maximum Clique Problem: from Classical algorithms to Graph Neural Networks and Quantum algorithms,
[\href{https://arxiv.org/abs/2403.09742}{https://arxiv.org/abs/2403.09742}].

\bibitem{Sloane-2000}
Neil J. A. Sloane, 2000,
\newblock Challenge problems: Independent sets in graphs,
[\href{https://oeis.org/A265032/a265032.html}{https://oeis.org/A265032/a265032.html}].


\bibitem{No-2019}
Albert No, 2019,
\newblock Nonasymptotic Upper Bounds on Binary Single Deletion Codes via Mixed Integer Linear Programming,
\textit{Entropy} \textbf{2019}, 21(12), 1202.


\bibitem{Nakasho-2023}
Kazuhisa Nakasho, Manabu Hagiwara, Austin Anderson, J. B. Nation, 2023,
\newblock The Tight Upper Bound for the Size of Single Deletion Error Correcting Codes in Dimension 11,
[\href{https://arxiv.org/abs/2309.14736}{https://arxiv.org/abs/2309.14736}].




\bibitem{Butenko-2009}
S. Butenko, P. M. Pardalos, I. Sergienko, V. Shylo and P. Stetsyuk, 2009,
\newblock Estimating the size of correcting codes using extremal graph problems,
In: \textit{Pearce, C., Hunt, E. (eds) Optimization: Structure and Applications}, pp. 227-243. Springer New York, 2009.


\bibitem{Sloane-2002}
N. J. A. Sloane, 2002,
\newblock On single-deletion-correcting codes,
in \textit{Codes and Designs, Ohio State University, May 2000 (Ray-Chaudhuri Festschrift)}, K. T. Arasu and A. Seress (editors), Walter de Gruyter, Berlin, 2002, pp. 273-291,
[\href{https://arxiv.org/abs/math/0207197}{https://arxiv.org/abs/math/0207197}].

\bibitem{VT-1965}
R. R. Varshamov and G. M. Tenengolts, 1965,
\newblock Codes which correct single asymmetric errors (in Russian),
\textit{Avtomatika i Telemekhanika}, 26 (No. 2, 1965), 288-292,
English translation in \textit{Automation and
Remote Control}, 26 (No. 2, 1965), 286-290.


\bibitem{Butenko-2002}
S. Butenko, P. M. Pardalos, I. Sergienko, V. Shylo and P. Stetsyuk, 2002,
\newblock Finding maximum independent sets in graphs arising from coding theory,
In: \textit{Proceedings of the 2002 ACM Symposium on Applied
Computing}, pp. 542-546 (2002).



\bibitem{Sloane-1989}
N. J. A. Sloane, 1989,
\newblock Unsolved Problems in graph theory arising from the study of codes,
\textit{Graph Theory Notes of New York}, 18 (1989), pp. 11-20.


\bibitem{Krpan-2024}
Alja\v{z} Krpan, Janez Povh, Dunja Pucher, 2024,
\newblock Quantum computing and the stable set problem,
[\href{https://arxiv.org/abs/2405.12845}{https://arxiv.org/abs/2405.12845}].


\end{thebibliography}
\end{document}